\documentclass[fleqn,twoside]{article}
\usepackage{espcrc2}
\usepackage{epsfig}

\usepackage{graphicx}
\usepackage[figuresright]{rotating}

\newcommand{\AmS}{{\protect\the\textfont2
  A\kern-.1667em\lower.5ex\hbox{M}\kern-.125emS}}

\title{The phase diagram of QCD with two and four flavors: results with HYP
       fermions
       \thanks{Talk presented by A. Hasenfratz}
      }

\author{A. Hasenfratz\address[MCSD]{Department of Physics, University of Colorado, 
        Campus Box 390, Boulder, CO 80309, USA}
        and
        F. Knechtli\addressmark[MCSD]
        \thanks{New address: Institut f{\"u}r Physik, 
        Humboldt-Universit{\"a}t zu Berlin, Invalidenstr. 110, 
        10115 Berlin, Germany}
       }

\begin{document}

\begin{abstract}
We study the finite temperature phase transition of four and two flavor
staggered fermions with hypercubic smeared link actions. These preliminary
studies suggest that the improved flavor symmetry of the fermionic action
can have a significant
effect on the finite temperature phase diagram.
\vspace{1pc}
\end{abstract}

\maketitle

\section{INTRODUCTION}
\vskip -0.0cm

In this talk we present preliminary results
of the two and four flavor QCD finite temperature phase diagram 
on \( N_{T}=4 \) temporal lattices using
a hypercubic blocked (HYP)\cite{coloquench},\cite{roland01}
 staggered fermion action. 
Because
of the large lattice spacing these simulations are more relevant to
illustrate the effect of the HYP action than to extract physical values.
Our goal here is to show how the improved
flavor symmetry of the HYP fermions influences the phase diagram.

In any fermionic action the thin link gauge
connections can be replaced by some sort of smeared links. Smearing partially
removes short range vacuum fluctuations (dislocations) improving the
topological properties of the gauge field configuration and improving
the chiral properties of the fermionic action. At the same time smearing
can distort the short range properties of the configuration introducing
new type of lattice artifacts. An optimal smearing is a transformation
where short range distortions are minimal while most of the dislocations
are removed. The hypercubic blocking was designed to be very compact,
and its  parameters are optimized non-perturbatively to remove most
of the vacuum fluctuations. Interestingly the non-perturbatively optimized
parameters remove the perturbative tree level flavor symmetry
violating terms of the action.

The HYP blocking can be combined with any fermionic action. Here we
consider staggered fermions. In \cite{coloquench} we showed 
in the quenched approximation that flavor symmetry
with HYP staggered fermions is improved by about an order of magnitude
relative to thin link staggered fermions. The improved flavor symmetry
is especially important for physical quantities that are sensitive
to chiral symmetry. 
The finite temperature phase transition
restores chiral symmetry at vanishing quark mass. Theoretical predictions 
\cite{P_W}
for this transition are based on the full chiral and flavor 
symmetry of the fermionic action. Lattice simulations studying 
the nature of the finite temperature phase transition at small 
quark masses could therefore  be very
sensitive to chiral symmetry violations of the lattice action. Since
the finite temperature phase transition occurs at large lattice spacing,
\( a\approx 0.3-0.15 \)fm on typical temporal lattices \( N_{T}=4-8 \),
the use of chirally improved actions could be essential there.

Before presenting our numerical results we mention
a puzzle about the flavor dependence of the finite temperature phase
transition. Phenomenological instanton model calculations predict
a strong flavor dependence for the chiral transition temperatures \cite{shuryak}.
For \( N_{f}=2 \) flavors the chiral transition is expected to be second order
at \( T_{c}\approx 150 \)MeV with crossover at finite quark mass
values. For \( N_{f}=3 \) flavors the transition is predicted to
be first order, \( T_{c}\approx 100 \)MeV in the chiral limit while
for \( N_{f}=4 \) flavors the chiral transition, if it exists at all,
is at a very small temperature. At finite quark mass both the \( N_{f}=3 \)
and 4 flavor cases show first order phase transition that persists
up to some critical quark mass value and the phase transition temperature
depends strongly on the quark mass.
Lattice simulations observe
only a weak flavor dependence, especially for \( N_{f}=4 \) flavors. Instanton
models rely on plausible but unproven assumptions yet they predict
the zero temperature chiral behavior of QCD very successfully. It
is puzzling why their prediction is  different from lattice calculations in 
the finite temperature case.

\section{\protect\( N_{F}=4\protect \) FLAVOR SIMULATIONS}
\vskip -0.0cm

We performed simulations on \( N_{T}=4 \) temporal
lattices with \( N_{S}=8,10 \) and 16 spatial size. The details of the simulations
can be found in \cite{proc_fr}. To see the effect
of smearing first we compare thin link, one level of APE and HYP smeared
actions at approximately matched quark masses. In quenched simulations
we found that the three actions are matched at quark masses \( am_{\rm HYP}=0.1, \)
\( am_{\rm APE1}=0.08 \), and \( am_{\rm thin}=0.06 \). Dynamical simulations
show that once the quark masses are matched, the lattice spacing at
fixed gauge coupling \( \beta  \) remains approximately matched also.
\begin{figure}
\epsfig{file=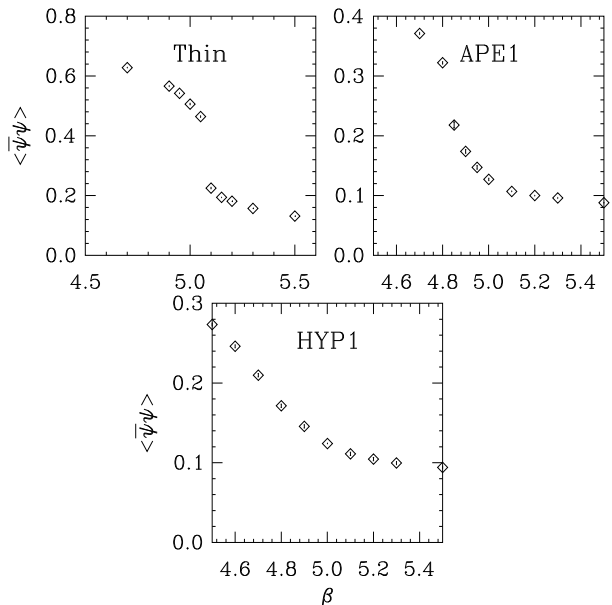,width=8cm,height=8cm}
\vskip -1.1cm
\caption{The chiral condensate \protect\( <\bar{\psi }\psi >\protect \)
for the thin link, APE1 and HYP actions at approximately matched quark
masses ($N_f=4$).
\vskip -0.8cm
\label{N1_pbp}}
\end{figure}
As it can be seen in figure \ref{N1_pbp} the strong
first order transition observed with thin link action weakens and
disappears as smearing is introduced at constant physical quark mass.
One might wonder if the HYP action might show a first order transition
at some smaller gauge coupling. We cannot exclude this possibility
without actual simulations, but even if there is a first order phase
transition, it occurs at a much smaller temperature than 
the phase transition of 
the thin link action suggesting that it is not a physically relevant
transition.

We did not find first order signals even at smaller
quark masses close to the physical light quark mass 
indicating that our simulations were all above the end
point of the first order transition line. To identify the crossover
it is traditional to look at the susceptibility of the quark condensate.
We found only very broad peaks in the susceptibility.
However the peak of the susceptibility
does not necessarily identify the crossover transition. It is more
sensitive to 
 nearby critical points like the end point of
the first order line. If the transition line has a strong quark mass-temperature
dependence, this could be very different from the actual crossover
region. Because of that concern we decided to use the chiral condensate
at fixed gauge coupling 
to identify the crossover temperature. At small quark masses we expect
the condensate to depend linearly on the quark mass,
{\[
\frac{<\bar{\psi }\psi >}{m_{q}}=\frac{\Sigma }{m_{q}}+c\]
 where \( \Sigma  \) and \( c \) are constants independent of the
quark mass. We distinguish three different cases.
If all quark masses  are in the chirally
symmetric phase, \( \Sigma =0 \). If  all quark masses   are
in the chirally broken phase, \( \Sigma >0 \) is the chiral condensate. 
If at fixed gauge coupling the smaller quark mass points are in the chirally 
symmetric while the larger quark mass points are in the chirally broken phase,
the formers predict $\Sigma=0$ while the latter ones 
predict a non-zero 
condensate at the transition point. \( \Sigma  \) however is not
this value but its linear extrapolation to zero quark mass and could
take any value, even negative ones. Figure \ref{pbp_extrapolate_nf4}
illustrates such an extrapolation at \( \beta =4.8\), 4.9 and 5.0.
The \( \beta =5.0 \) points are all in the chirally symmetric phase,
the other two are mixed. From the plot we can
read off the transition mass at fixed gauge coupling. The negative
slope at larger quark masses indicates that the condensate at the crossover
is small.
\begin{figure}
\epsfig{file=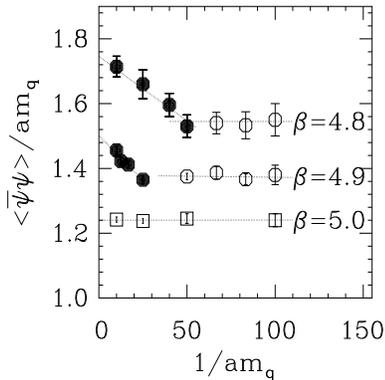,width=5cm,height=5cm}
\vskip -1.1cm
\caption{\protect\( <\bar{\psi }\psi >/am_{q}\protect \) as the function
of \protect\(1/am_{q}\protect \) at three different gauge coupling
values ($N_f=4$).\label{pbp_extrapolate_nf4}}
\vskip -0.8cm
\end{figure}
The predicted crossover region is plotted in figure
\ref{mt_vs_beta}. At \( \beta =5.0 \) the crossover occurs above
\( am_{q}=0.1 \) corresponding to a temperature around \( 220 \)MeV.
At \( \beta =4.95, \) \( am_{t}=0.04 \) the crossover temperature
is approximately \( T=170 \)MeV. Whether this temperature changes
significantly as we move to smaller quark masses has to be investigated
in the future.
\begin{figure}
\epsfig{file=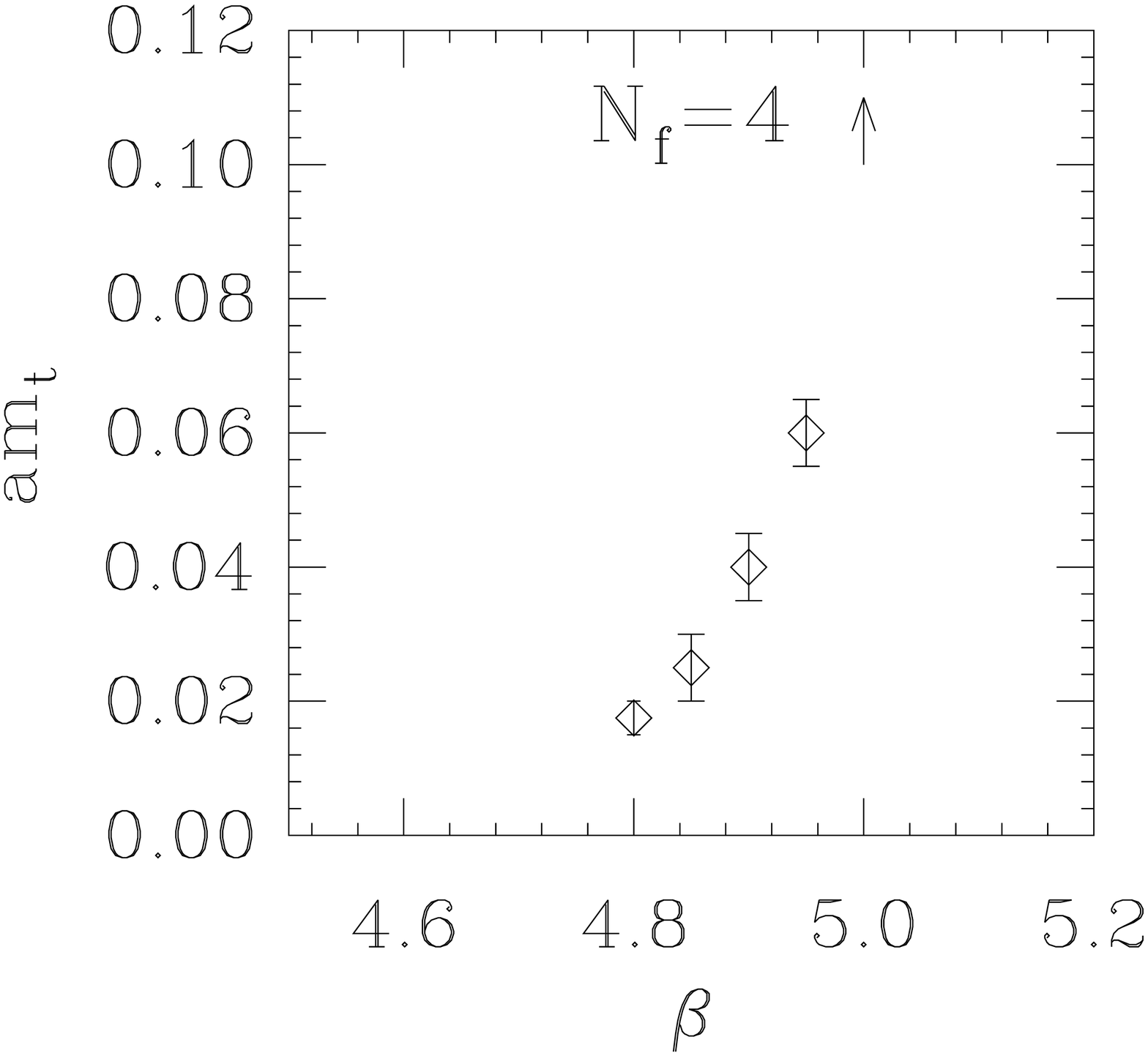,width=5cm,height=5cm}
\vskip -1.1cm
\caption{The predicted crossover line for the \protect\( N_{f}=4\protect \)
flavor system on \protect\( N_{T}=4\protect \) temporal lattices.
\vskip -0.8cm
\label{mt_vs_beta}}
\end{figure}
In any case our data indicate that the first order end point occurs
at a fairly small quark mass value (and possibly small temperatures). 

\section{\protect\( N_{f}=2\protect \) FLAVOR SIMULATIONS}
\vskip -0.0cm

We have preliminary results
  of the two flavor system on \( 8^{3}\times 4 \) lattices
at several quark masses. In the simulations 
we approximate the square root of the fermionic
determinant with a finite polynomial. 
Some of the details of the algorithm are discussed in \cite{proc_fr}. 
The phase diagram does not show significant
deviation from the thin link action results. Figure \ref{pbp_nf2}
shows \( <\bar{\psi }\psi >/am_{q} \) as the function of the gauge
coupling for \( am_{q}=0.01 \) and \( am_{q}=0.04 \).
The figure suggests a phase transition around \( \beta =5.2 \).
Further data at quark mass \( am_{q}=0.02,0.03 \) and 0.06 support
this conclusion. The phase transition for the two flavor system on
\( N_{T}=4 \) is a crossover, it occurs around \( \beta =5.2 \)
and depends only weekly on the quark mass. The susceptibility of the
chiral condensate predicts the same value. 
\begin{figure}
\epsfig{file=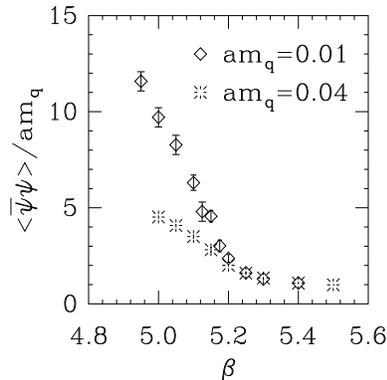,width=5cm,height=5cm}
\vskip -1.1cm
\caption{\protect\( <\bar{\psi }\psi >/am_{q}\protect \) as the function
of the gauge coupling for \protect\( N_{f}=2\protect \) flavors at
two different quark mass values.\label{pbp_nf2}}
\vskip -0.8cm
\end{figure}

\end{document}